# Quantum Hall Effect at 40 kelvin:
## Evidence of Macroscopic Qantization in the Extreme Soft Limit

Timir Datta, Michael Bleiweiss, Anca Lungu and Ming Yin

Physics & Astronomy Dept., and the Nanocenter
University of South Carolina, Columbia, SC 29208, USA.

Zafar Iqbal
Chemistry Department
New Jersey Institute of Technology
Newark, NJ, USA.

Quantum Hall Effects [1,2] (QHE) are of two types, [3,4] the integer effect is a direct result of Landau level quantization of noninteracting electrons, and the fractional effect similar to superconductivity, is a macroscopic quantum phenomenon but arise from a new form of electronic condensation into a strongly correlated quantum liquid (QL). Usually QHE are observed in very high mobility, two-dimensional, electron (hole)-gas or (TDEG) systems typically under high magnetic fields (B) and at low temperatures (T), i.e., in the extreme quantum limit (B/T>1). Quantum Hall effect is applied as calibration benchmark, international resistance standard, and a characterization technique for semiconductor heterostructures. Applications can be widespread if the devices and the operating conditions were more accessible. Here we report evidence of both fractional and integer quantum hall effects (2/3, 4/5, 1, etc) in a novel carbon structure, in a remarkably soft quantum limit of T~ 40K.

As indicated above, the devices that contain the TDEG themselves are complex quantum well structures. The fabrication of these hetero-junctions is a marvel of modern semiconductor manufacturing. Not surprisingly the exploration for new simpler systems and understanding the nature of the quantum hall states under low magnetic fields[9] are contemporary topics of research.

Our strategy to achieve the desired quantum properties has been by modifying the materials geometry and the mesoscale confirmation. Mesoscale shape[10] and nontrivial atomic coordination[11-14] are known to enhance quantum–mechanical transport properties. The system of choice belongs to a new class of solids, replica opals, three-dimensional carbon replica opals (CRO) in particular. Introduced by Bogomolov et al[15], these are structurally complimentary to artificial (porous) opals and have recently received a considerable attention as a new source of self-aggregated, nano-structured solids. The properties of these materials depend on the opal sphere size and preparation. They form optical cavities[16] photonic bandgap crystals[17] and superconductors[18].

Our specimens were fabricated from silica opals, by a process of chemical vapor deposition (CVD) of carbon from propylene ($C_3H_6$) gas; carbon grows as surface layers on the silica spheres to form a closed packed stacking of mesoscopic "carbon-eggshells". After the desired amount of carbon is deposited the silica spheres are removed by chemical etching. More details about deposition may be found in reference 16. The freestanding carbon regions are multiply connected[19] and form a lattice of interconnected percolating paths. The thickness of the carbon is uneven and depends on the (local) infiltration rate of the gas. For example no deposition takes place so thickness is zero and holes are left behind at places where the vapor penetration is obstructed, such as in the contact regions between the silica spheres.

The sphere diameter of the samples is about 250nm. A picture of a cleaved surface is shown in figure 1. The inset at the left corner is a color-enhanced image of the interior of two half shells. The one on the lower right is another view of the shell stacking. The arrows mark some of the foramens in the carbon shells.

For transport experiments the specimens were cut into small rectangular (2x4x5mm) blocks and gold or copper electrodes were attached in the standard four and six terminal geometry. Phase sensitive measurements were performed in first our laboratory at USC, for temperatures upto 40K and at low fields (B<3 tesla) in an ultra-low noise, cryostat with a superconducting magnet. Through out a run T held constant, and B was varied in steps according to a preset program. Each measurement was taken with B held constant and the magnet in the "latch" mode. At the National High Magnetic Field Laboratory, (NHMFL) in Tallahassee, Florida, higher field (B<20 T) experiments were performed while B was ramped up and down.

We observe that these samples were electrically non-metallic, i.e., resistance, R(T), increases as temperature was reduced (figure 2). However, charge transport is not simply activated at low temperatures evidence for weak localization[19] and carrier motion was via variable range hopping (VRH) between localized states were observed [20]. The inset plot of figure 2 shows localization and VRH behavior.

With a magnetic field (B) applied along the z-axis and current flowing along x-axis both the longitudinal magneto response ($R_{xx}$) and that perpendicular to both current flow and B i.e., Hall voltage, $V_{xy}$ (or $R_{xy}$) were measured simultaneously. As B was increased hall resistance followed the well-known signature of quantum hall effect - a unique step pattern. Correlated with the steps of $V_{xy}$, $R_{xx}$ showed quantum-oscillations due to Subnikov-deHass (SdHass) effect. Figure 3 is a von Klitzing plot of the $V_{xy}$ and $R_{xx}$ isotherms at 38K. Even at this relatively elevated temperature Subnikov oscillations and hall plateaus were resolved with good synchrony between $V_{xy}$ and the magneto-resistance.

It may be shown[21] that a direct proportionality exists between the field B of the SdHaas minima and $V_{xy}$, or the corresponding resistance $R(\nu)_{xy}$ steps as follows,

$$B = (en_s)R_{xy} \qquad \ldots 1$$

Where, e is the unit of electric charge and $n_s$ is the carrier density and $n_s$ can be determined from a fit of the data to equation 1, and the values of the Landau-levels filling factor ($\nu$) corresponding to the plateaus in the hall curve are given by §:

$$(\nu)^{-1} = (\frac{e}{n_s h})B \qquad \ldots 2$$

The hall plateaus in figure 3, were designated by the values of $\nu$ as estimated from equation 2.

The Landau-levels that were particularly prominent, are the rational fraction filling factors $\nu = p/q$ (p,q are integers) 2/3, 4/5 and the integers 1 and 2 another high integer step with fill factor of 12 was also noticed. It may be useful to compare some literature values of field and temperature required for QHE in high mobility, samples; for example in a 500Å wide modulation-doped GaAs/AlGaAs, heterostructure in 35 mK, the "2/3" state is observed at a little over 6 tesla[22]. In $Bi_{2-x}Sn_xTe_3$ the "2" state is recorded at 0.3 K and ~9 T[23]. High filling factor states can be condensed at low magnetic fields, such as ~0.2 tesla for the "36" state, but at a temperature of 50mK[24]. These are orders of magnitude higher than those required for the carbon replica opal samples (B/T ~ 0.7/38~1/50) presented here.

Dividing the right hand side of equation 1 by the quantum unit of resistance ($h/e^2$) and the left hand side by B(1), the value of the magnetic field at the SdHass minimum associated with the unit hall resistance step then we obtain:

$$\frac{B}{B(1)} = \frac{R_{xy}(\Omega)}{(h/e^2)}$$

or in these normalized units of field ($B_{norm}$) and resistance (R*),

$$B_{norm} = R^* \qquad \ldots 4$$

Equation 4 predicts that any set of QHE data when plotted in the variables $B_{norm}$ and R*, follow a direct universal direct proportionality with unit slope. This proportionality holds irrespective of the (positive or negative) sign of the carrier or their density and independent of the sample temperature. To test this proportionality we normalized the $R_{xy}$ and B data of figure 3 by dividing with 25812.4 Ω and 1.44 T respectively. Also for the purposes of comparison we digitized some QHE-literature values; such as, the data for ν equal to 2,3,4,6 & 8 states in a crystal of $Bi_{2-x}Sn_xTe_3$[22]; and for the 14,16,18, 20 & 22 states[24] in a sample comprising of lateral superlattices. Figure 4 provides a graphical demonstration of the universal unit proportionality of all the $B_{norm}$ and R* quantum hall effect in these systems [25-27] including those of figure 3.

Quantum hall effect is a proof of charge carrier condensation in two-dimensional (2D) states. The sheet density ($n_s$) of the carrier in QHE systems vary considerably, in typical heterostructure [3,4,21] $n_s$ is $\sim 10^{15}$ m$^{-2}$. For our system, we determine $n_s = 2.2 \times 10^{14}$ m$^{-2}$ from the slope of the B vs $R_{xy}$ plot (equation 1). So these carbon replica opals are a low carrier density 2DEG system comparable to that of reference 26.

In bulk systems QHE is possible [23,27-30]. For example, Belenkii has reported integer effects in the defect states of layered bulk semiconductors such as InSe. Quantization steps associated with field induced spin-density-waves were reported in the Bechgaard salt $(TMTSF)_2PF_6$ and Kul'bachinskii observed integer QHE in $Bi_2Te_{23}$, $Sb_2Te_{23}$ and $Bi_{2-x}Sb_xTe_3$.

Carbon replica opal is a material with a rich variety of the low-dimensional features and an enormously large ratio of surface to volume. It is possible that in the carbon system, the QHE arise from 2-d, interface or surface states. Surface states are importance for the condensation of QHE. Furthermore, in a bulk material low dimensional states can play a role[29] similar to the Anderson localized states for standard two-dimensional quantum hall systems. A central feature of the many-body, wave function that describes fractional quantum hall effect [3,4] is nodes. Having deep nodes in the wave function is one of the prime requirements of the Laughlin theory[30]. The holes or "anti dots" in the carbon matrix keep the carriers away from these void regions forcing nodes to form in their wave function. The three-dimensional, multiply connected, topology can provide many pathways to amplify quantum interference. Transformation of compressible and incompressible quantum fluids forming dots and anti-dots are important[31] in quantum hall systems.

In the carbon system, there are also self-similarity and geometric hierarchies such as "hole inside a hole". The importance of hierarchies in QHE has been a focus of current research attention[32-34]. Transport coefficients can be greatly affected by geometric effects. For example, in narrow gap, inhomogeneous semiconductors this effect may be so strong that SdHaas signals are detectable even at room temperature[10]. The combination of a plethora of geometric features and low carrier density may be favorable for the formation of quantum-hall condensates in the low B/T limit.

To the best of our knowledge, the current work is the first observation of quantum hall signals at temperatures well above the boiling point of liquid helium and in the very soft quantum limit. It is possible that QHE may occur at even higher temperatures. High temperatures will render these effects more accessible and may enable widespread technological applications.

**Acknowledgements:**

We wish to acknowledge Dr. Ray Baughmann for providing the specimens of carbon replica opal, Dr. Eric Palm for assistance with the high β facility at the NHMFL. The USC NanoCenter and the Materials Science and Technology Division of the Naval Research Laboratory provided partial supports to the first author.

**References & foot notes:**

§ Mathematically equations 2 is equivalent to the following equation:

$$(B)^{-1} = (\frac{e}{n_s h})\nu \qquad (2b)$$

But one may be preferable over the other. Typically, the form that distributes the ordinates and abscissas of the data points more evenly is chosen.

**Figures:**

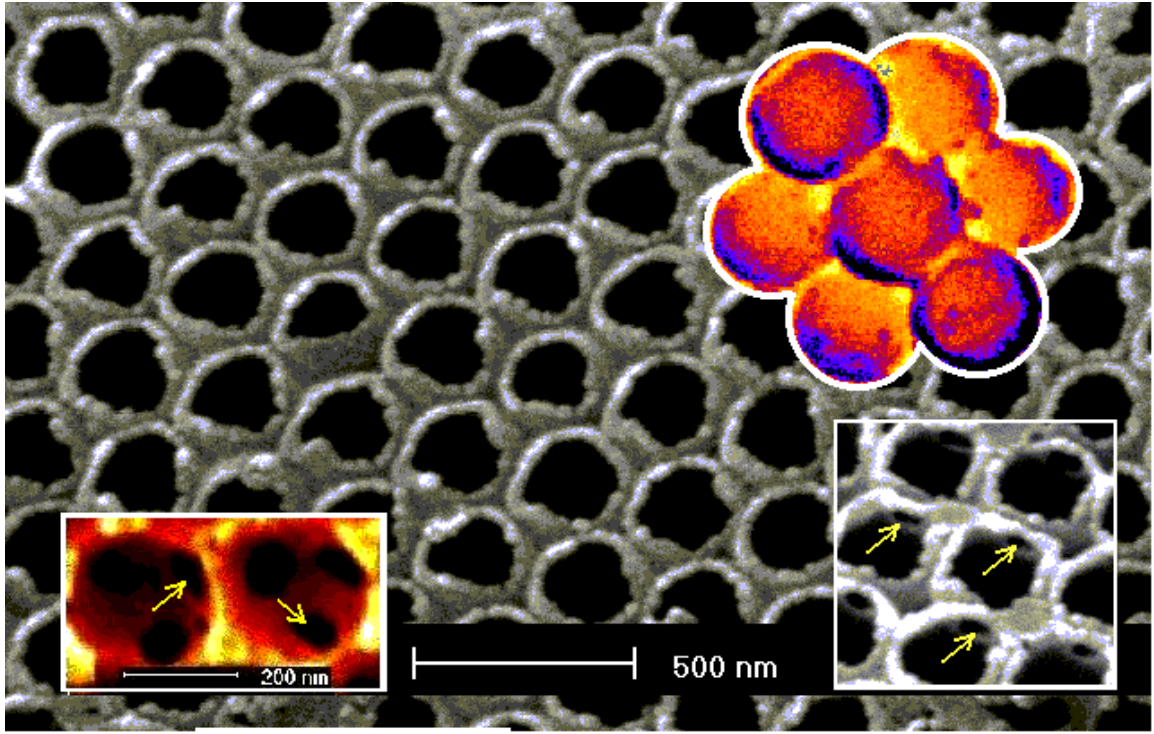

**Figure-1:** Electron micrograph of a QHE active, carbon replica opal (CRO) sample. The packing of the chemical vapor deposited carbon regions is shown. To show how the original opal underlies the replica structure, a color rendered image of the basic building block, a cluster of silica spheres is superposed on the top right hand corner. The insets at the bottom show details of the holes in the carbon. The size scales of the whole picture and that of the small inset are indicated.

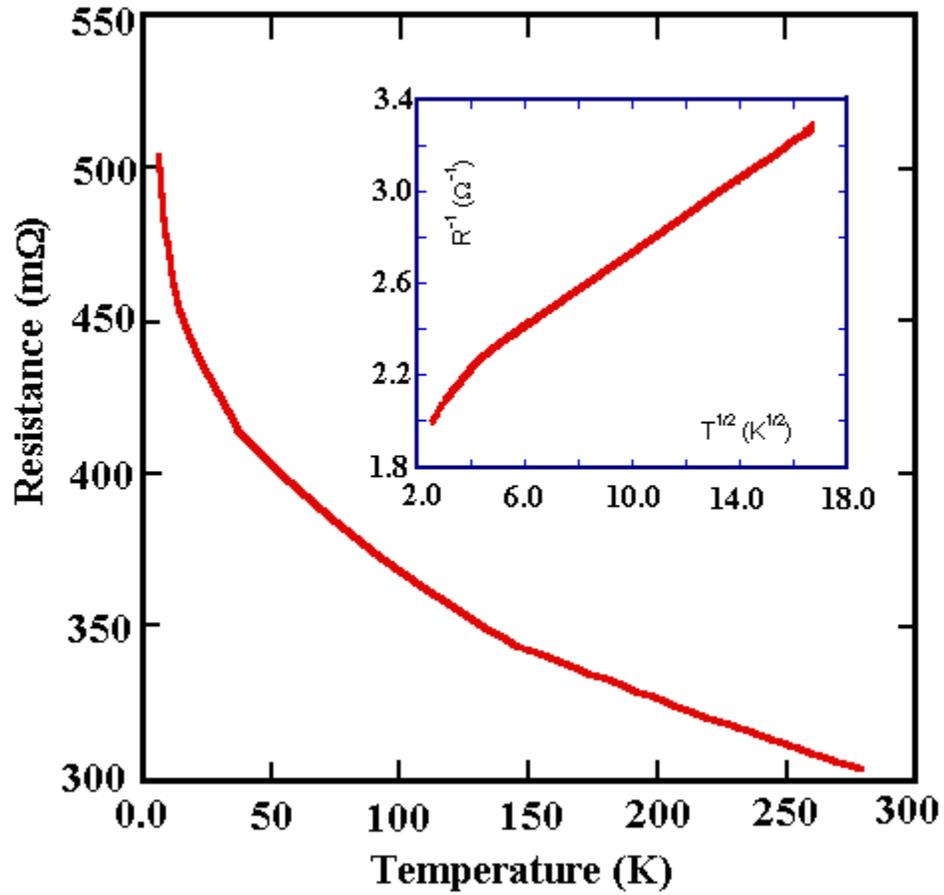

**Figure-2:** Electrical resistance of a carbon replica opal as a function of temperature. As may be seen the behavior is non metallic that is resistance increases as temperature decreases. Furthermore at low temperatures hopping conduction and weak localization are indicated as shown by the inset plot.

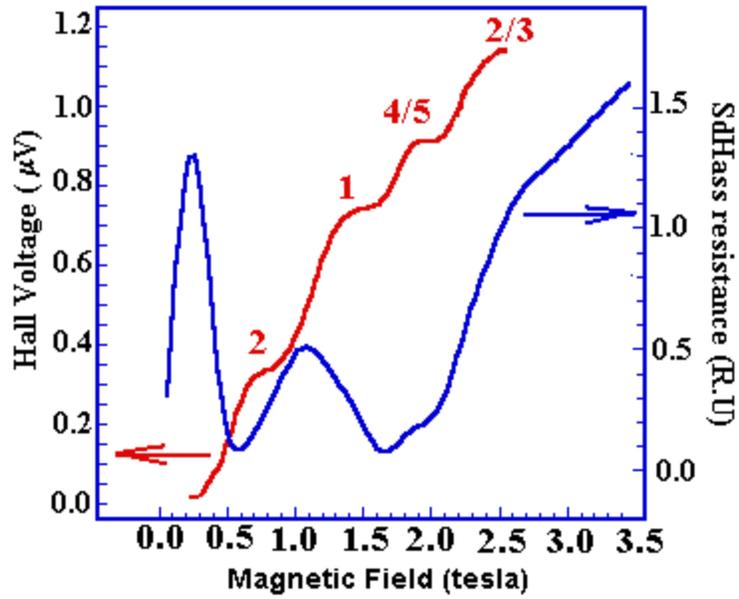

**Figure-3:** A von Klitzing plot at a temperature of 38 K shows the Shubnikov-deHaas oscillations (blue curve) and associated hall plateaus (red curve). The values of the filling factors estimated from equation 2 are also indicated. The arrows point the respective scales appropriate for each of the curves.

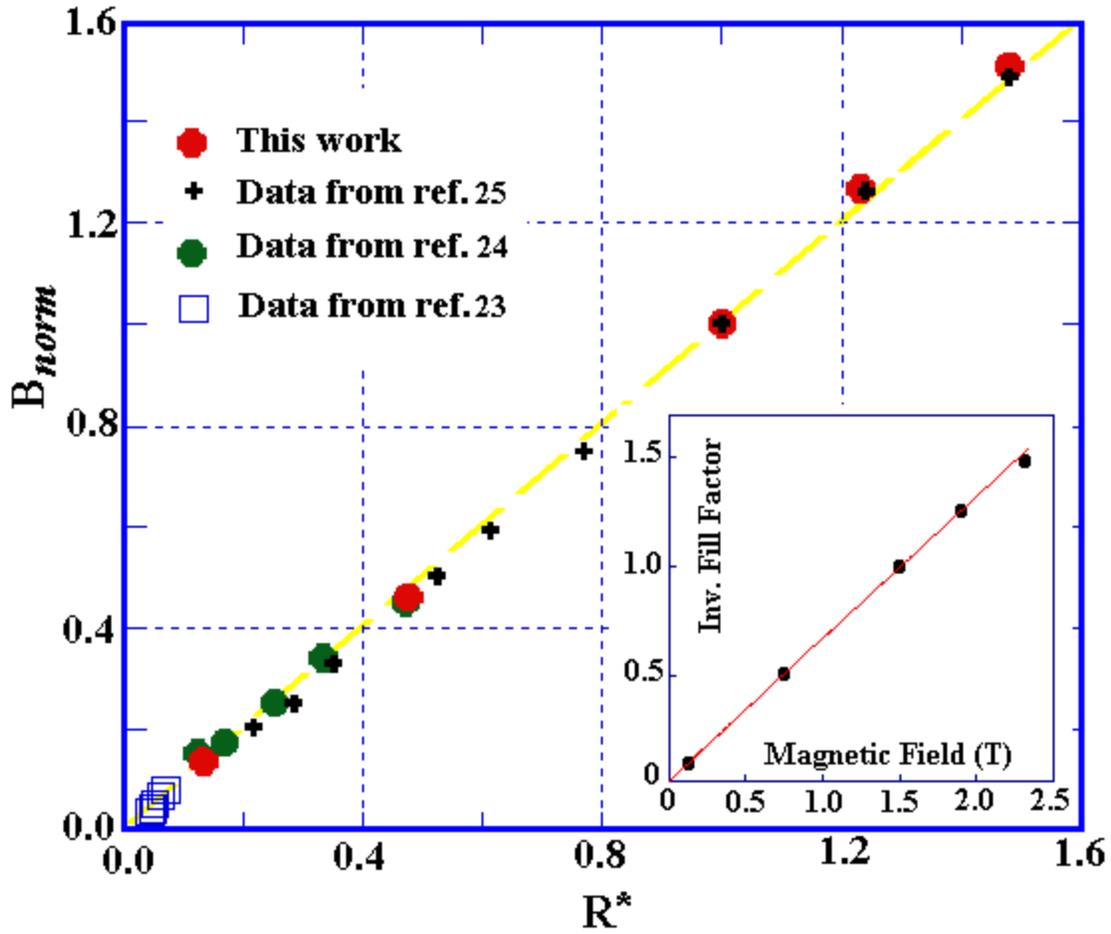

**Figure-4:** Universal proportionality of $B_{norm}$ (the magnetic field B normalized by the field at unit step) versus $R^*$, the value of resistance (in units of $h/e^2$) and comparison of quantum Hall effect data from this work and three other sources in the published literature [24-27]. The inset shows the fit of equation 2 used for the determination of the fill factor ($\nu$) values indicated in figure 3.